# Lightweight Service Oriented Architecture for Pervasive Computing

**Jean-Yves TIGLI[1,*], Stéphane LAVIROTTE[1], Gaëtan REY[1], Vincent HOURDIN[1,2] and Michel RIVEILL[1]**

**[1] I3S, University of Nice – Sophia Antipolis**
**Sophia-Antipolis, France**

**[2] MobileGov**
**Sophia-Antipolis, France**

## Abstract

Pervasive computing appears like a new computing era based on networks of objects and devices evolving in a real world, radically different from distributed computing, based on networks of computers and data storages. Contrary to most context-aware approaches, we work on the assumption that pervasive software must be able to deal with a dynamic software environment before processing contextual data. After demonstrating that SOA (Service oriented Architecture) and its numerous principles are well adapted for pervasive computing, we present our extended SOA model for pervasive computing, called Service Lightweight Component Architecture (SLCA). SLCA presents various additional principles to meet completely pervasive software constraints: software infrastructure based on services for devices, local orchestrations based on lightweight component architecture and finally encapsulation of those orchestrations into composite services to address distributed composition of services. We present a sample application of the overall approach as well as some relevant measures about SLCA performances.[1]

***Key words:*** *Software Composition, Pervasive Computing, Service oriented Architecture, Service for Device*

## 1. Introduction

Pervasive computing is omnipresent computers [1] in the real environment through a large number of objects and new devices in our everyday life (everyware [2]). Indeed, with the miniaturization of computer hardware, processing units become invisible and integrate into buildings, clothes, vehicles, and so on. They can be at the same time mobile, integrated and often coupled to the physical environment [3]. They increase application fields of computing by a growing quantity and diversity of smart devices in the physical environment of users [4]. For all these reasons, pervasive and ubiquitous computing appears like a new computing era [5] based on networks of objects and devices evolving in a real world, radically different from distributed computing, based on networks of computers and data storages.

In this paper we focus on the pervasive software challenges.

A classical way to address such topic is to consider dependencies between the real world and the software application in the so-called context-aware approaches. Lots of papers propose various approaches to take into account contextual information into the applications using various architectures ([6], [7], [8], [9]). We argue that is only a first step (or rather the last!) to take into account the dynamic evolution of the surrounding physical environment. The first real challenge is to find all the ways to interact with and recover such contextual information from the physical environment. This is classically the role of input/output devices in our computers. But this classical approach based on layered software provides standard runtime and libraries. It is based on predefined set of devices without being able to integrate numerous devices or objects on the fly, without a priori knowing them. In other words, Pervasive computing must deal with a dynamic software environment (called software infrastructure afterward), before processing contextual data.

The challenge we address in this paper is to propose a middleware for pervasive computing being able to deal with numerous objects and devices: being able to adapt to their intrinsic heterogeneity according to the used technologies, their behaviors and functionalities, being able to react to their appearance and disappearance at runtime.

We first study how existing middlewares for pervasive computing are taking into account these specific

---

[1] This work is currently supported by ANR project ANR-08-VERS-005

* Currently delegated as INRIA researcher in the team PULSAR

**IJCSI**

constrains, and we confirm that service oriented paradigm can be an efficient approach to meet some pervasive computing challenges. We then introduce some specific extensions before proposing the SLCA (Service Lightweight Component Architecture) model, as an original service oriented approach for pervasive computing. Then we present experimentations based on SLCA and we analyze some measurements and results. Finally we conclude on the limitations of such an approach and we introduce future works.

## 2. When SOA Meets Pervasive Computing

Among the numerous software paradigms, Service Oriented Architectures (SOA) [10] originally brought eight principles and influences to software engineering: service encapsulation, allowing any software to be run as a part of an architecture entity; loose coupling, minimizing the dependencies between services, and increasing dynamicity and reusability; service contract, forcing services to adhere to a communication agreement, providing descriptions of what they provide or require; abstraction, also referred as black-box abstraction, limiting knowledge of the service logic to the contract; service autonomy, making services independent from any other, and self-sufficient for the service they offer; service discoverability, allowing services to be dynamically discovered at run-time, with some criteria; finally composability, coordinating services and assembling them into composite services. Then considering objects and devices of software infrastructure as services, SOA is well-suited to deal with the heterogeneity and the dynamic evolution of pervasive systems as we defined in the previous section. Thus using SOA to create applications based on physical or virtual devices proved its worth for almost ten years, with Jini (1999) [11] and UPnP (1999) [12], then more recently DPWS (2004) [13]. Like services, devices are autonomous, independent, and provide a set of functionalities, which can be contractualized. We then talk about service as the basic entity of the environment, which can be a software service, or a service representing a device.

In such evolution, three new principles and features appeared to fit pervasive application constrains. Firstly, one of major evolutions is probably the full discoverability. In fact services are classically discoverable using repositories and service brokers. Discoverability can be interpreted as several ways to discover or use services of the environment. For example, it can mean that services are discovered at run-time, but only from a list of already known services like in Gaia [14]. Enterprise-oriented SOA models like CORBA [15] and OSGi [16] use centralized services registries in which service producers have to register themselves in order to be used by service consumers, like the UDDI registry for Web services. What we are interested in for discoverability is the fully dynamic, reactive, and decentralized discovery of services, as introduced by services for devices like Jini [6], SLP, or Web Services for Devices (WS4D) like UPnP (SSDP) or DPWS (WS-Discovery). In the case of pervasive applications, each entity must be able to discover locally all the services for devices in its context and to detect their appearance and disappearance dynamically [17]. This is a hot research topic in SOA for pervasive computing [18], [19], [20].

Secondly, a new kind of interaction between services and then devices is required to allow spontaneous messages to be sent from the real world, from objects and devices, to software applications. The new event communication feature suggested by Event-Driven Architectures (EDA), in which is a part of Advanced SOA, adds such required reactivity to pervasive applications. Most pervasive computing middlewares (Amigo [21], Aura [22], Gaia [14], Oxygen [23]) use standard remote method invocation (RMI) or remote procedure calls (RPC) technologies to interact with devices. With the outcome of EDA in Advanced SOA and with WS4D, eventing communications have been integrated to SOA to decrease coupling between entities and to increase reactivity of applications and systems. The CORTEX middleware for pervasive computing bases itself on a publish/subscribe event management for message passing between all its sensors and actuators.

At last, the interoperability requires making interactions between services and devices independent of the communication technologies used. The most popular approach in this field is Web service technology, adding Web standards based protocols which is the case in WS4D, for example, in Amigo. In Aura however, a specific XML description format is used, and connectors to services are created depending on the communication protocol used (CORBA, COM, or RPC), thus maintaining an interoperability layer without the use of Web services.

Table 1: Main characteristics of major works in pervasive computing

|                  | Amigo | Aura | CORTEX | Daidalos | Gaia | Oxygen |
|------------------|-------|------|--------|----------|------|--------|
| Discoverability  | X     | X    | X      | X        |      | X      |
| Reactivity       |       | X    | X      |          | X    |        |
| Interoperability | X     | X    |        | X        |      |        |
| Composability    | X     |      |        | X        |      |        |

X for the main principles handled by the different projects

At this stage, the main remaining challenge is the way to dynamically compose such services into an overall efficient and valid application. Despite advantages of service composition in ambient computing, for example easing discovery of relevant services in the current context or providing a set of services fitting users needs [24], only a few middlewares handle composability.

Indeed, among the above cited projects, only Amigo supports dynamic and context-sensitive service composition, bringing new pertinent services to the environment. We can however cite the Daidalos project [24], a middleware for composition of pervasive mobile services, which handles decentralized dynamic discovery and service composition, but which does not handle evented communications. The Table 1 summarizes how the main principles are handled by the different projects we have seen. When discoverability is checked, it means that the project supports full discoverability. Reactivity denotes the use of evented communications. Interoperability means that non-language-dependent representation is used for service description, and that entities from multiple programming languages and operating systems can interact in the same middleware. Finally, composability refers to the ability to export new entities of the same type in the infrastructure, using other discovered entities.

In conclusion, from what we have studied, there is no middleware tackling all the principles of pervasive computing based on event-based service architectures.

In the next section we introduce our original approach called SLCA to dynamically compose numerous services for devices and objects. First we explain how we can consider everything as a service. Then we propose a first way to locally orchestrate various services using a lightweight component based approach. Finally we present how we can reuse such local orchestration as new services in an overall distributed composition of services for real pervasive applications.

# 3. Our SLCA Model

SLCA (Service Lightweight Component Architecture) is a model of architecture for event-based service composition based on an assembly of lightweight components. The SLCA model relies on a software and hardware execution environment evolving dynamically. We define this environment as a set of resources, which are software or hardware entities provided to the application and that can appear and disappear at runtime.

Following the reasons mentioned in previous chapters, we propose an architecture taking into account three main paradigms:

– Web service oriented architecture. Ambient computing applications are then a graph of Web services and composite Web services. Interoperability, distribution, and discoverability are then assured.

– Lightweight assembly of component. Composite Web services are created from a dynamic assembly of black box components, executing in a local container, which doesn't provides mandatory technical services (non-functional concerns). Dynamicity of applications is thus provided, and reusability is increased.

– Events. They take place in the model at the service level, with Web services for devices for example, as well as in lightweight assemblies of components. Their advantages are twofold: they promote reactivity of systems, and increase decoupling between entities, and thus dynamicity of applications.

SLCA thus defines a compositional architecture model based on events, to design composite Web services, and increment the cooperation graph of services and applications. The environment consists of mobile users interacting with the world or users with worn or mobile devices. We see them as services momentarily available in the infrastructure. Composite services use services of the infrastructure as required interface to create new applications or to add new functionalities and export them on their new provided interface.

In the next subsections, we will explain more deeply what constitutes the SLCA model, and illustrate all of the three points, that are the service infrastructure, the service orchestration, and the service composition.

## 3.1 Pervasive Software Infrastructure of Services

SLCA is based on a service infrastructure using events, and dynamically discoverable in a decentralized way. They represent devices used in ambient computing applications, as well as composite services created by SLCA. Interoperability is maximal, thanks to the use of Web services, which can be used or implemented with any programming language and on all hardware architectures.

The architecture is completely dynamic. Services appear and quit the network reflecting the presence of devices, without knowing beforehand any service registry. It is possible to take into account these changes in applications without knowing what devices shall be met at design-time. Indeed, from the XML description of Web services, it is possible to automatically generate proxy components which will enable communications with services of the software infrastructure.

The service infrastructure of SLCA architecture is thus used for the discovery and the communication with services distributed in the environment. Applications are designed by dynamic service orchestration (mashups).

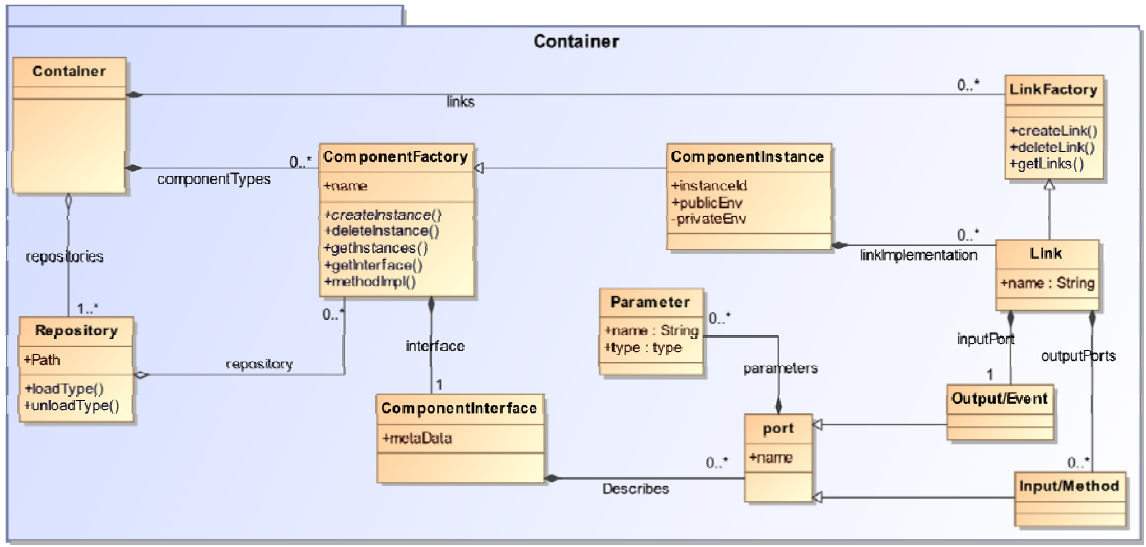

Fig. 1 LCA meta-model: lightweight components

An example of such a service infrastructure would be the set of services that a room can contain. All devices that are present inside this room are able to provide a service interface, either natively or by bridging native protocol to a service layer: lights, shutters, air-conditioning, TV, sensors, video projector… A user entering this room is then able to discover all these devices, and communicate with them in an homogeneous way, using advanced Web services.

This is the service infrastructure of SLCA. We will now focus on how service orchestration is made, to create actual applications. For that matter, we use lightweight service composition in our model.

### 3.2 Lightweight Services Composition

SLCA composition approach is based on lightweight components [25] similar to JavaBeans and OpenCom [26], but to design Web services orchestration. A composite service encapsulates the SLCA container which contains a dynamic lightweight component assembly. The LCA (Lightweight Component Architecture) component model is a model derived from Beans [27], adapted to other programming languages, with concepts of input, output ports and properties.

Like in most lightweight component based approaches [26], these components are called 'light' for several reasons. The first is that they execute in the same memory addressing space, and in the same process. Their interactions are thus reduced to the simplest and the more efficient way: the function call. The second reason, which stems from the first, is that they don't embed non-functional code for middleware or other non-mandatory technical service in this local environment. Their memory footprint is then reduced and they are instantiable and destructible quickly. To finish, they don't contain any reference between them at design-time, and respect black boxes and late-binding concepts. The dynamicity of the model is thus maximal, since they use events to communicate between them; components are fully decoupled, and highly reactive.

The only non-functional code present in the components is event management and properties accessing. Higher level programming languages define these operations; component code is then a simple object, like JavaBeans or .NET components, not overloaded with code injection for any purpose. The container does not provide technical services easing the programmer work, but consequently allows the creation of components with various requirements, like components needing to access hardware and thus low level functions. Adding non-functional properties, like security, journaling, or persistence of messages can be made by adding components in the assembly, guaranteeing scalability of the model.

As described in the LCA model (Fig. 1), components have an interface, defined by the component's type. This interface is a set of input ports (methods), and output ports (events), each one being typed by its parameters, and having an unique identifier. Interactions between components are bindings or links.

They link an output port of a component to one or more input port of components. Ports being explicit, no code has to be generated, nor studied by introspection to know what to modify in components to change the target of a binding at run-time. When an event is emitted, the control flow is passed to recipients in an undefined order, but this can be fixed adding sequence components. When limiting to unique bindings, and using sequence components, control flow managing of the application is fully deterministic. Not having indirections, due to technical services of the framework, gives a full control on control flows, and eases their debugging.

Component types which can be instantiated in a container depend on the list describing and implementing them in a repository. This list is also modifiable at run-time. When a service is discovered, its corresponding proxy component can be immediately loaded and instantiated in the component assembly to contribute to functionalities of the composite service.

Component assemblies inside composite services can create applications or new functionalities from services of the environment. Unlike the service infrastructure, they are executed locally, and their logic is not disturbed by appearing or vanishing of services.

When a service used by a composite service becomes unavailable, there are two possible reactions: either the state of the assembly is unmodified until a replacing compatible service is found, or its proxy component can be removed and the composite service can be adapted. In the first case, the locality of the assembly of component makes it able to save its state. Of course, adaptation mechanisms should be applied to take into account new requests to the composite service, which may or may not be able to completely satisfy a request. Continuing our room example, we climb here at the level of application creation. Indeed, we already have a service infrastructure, and we are now creating service orchestrations with lightweight component composition. These compositions will create dynamic applications, based on available services of the environment. For example, a user entering a room will be able to use all devices of the room to create his new application. If some devices are appearing in the room, because they are moved in by another user, or simply turned on, they shall be added to the current service based application.

## 3.3 Distributed Composition of Services

SLCA defines an architecture of composite event-based Web services, which are constructed by assembling components (Fig. 2).

A composite service then contains an assembly of components, in a container. Proxy components to other Web services are instantiated in the container of a composite service, and create applications from services present in the environment. Moreover, applications created through such orchestrations are exported as services to the service infrastructure (Fig. 3).

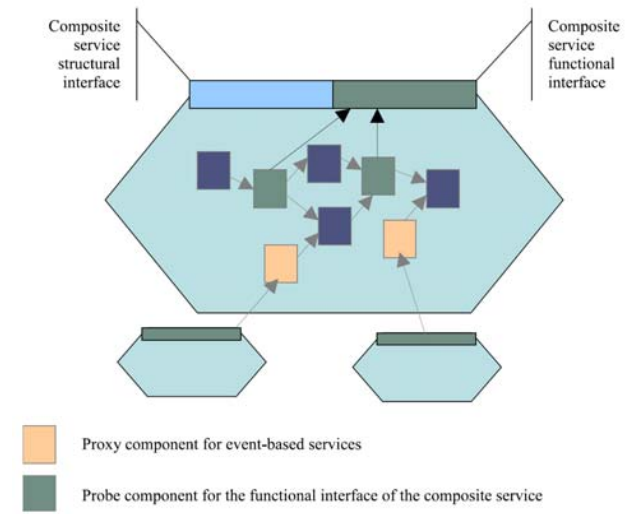

Fig. 2 Composite event-based Web service.

A composite service (container) provides two service interfaces (Fig. 2). The first one, the dynamic functional interface, allows publishing and accessing functionalities provided by the composite Web service; the second one, the control interface, allows dynamic modifications of the internal component assembly which provides these new functionalities.

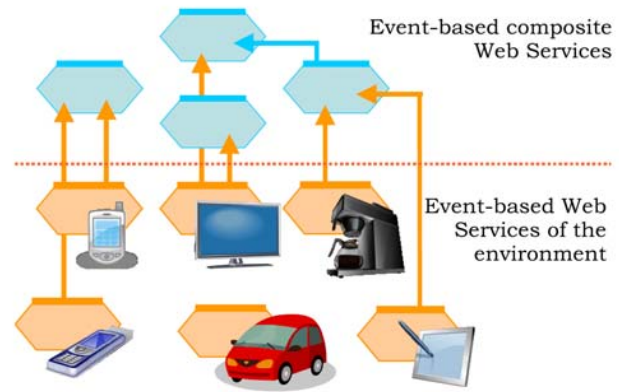

Fig. 3 Graph of event-based Web services.

The dynamic functional interface exports events and methods of the internal component assembly using probe components. Adding or removing a probe component dynamically modifies the functional interface and its description in the corresponding composite service. Adaptation to environment variations can be made by modifying the interface of a composite service, without stopping its execution.

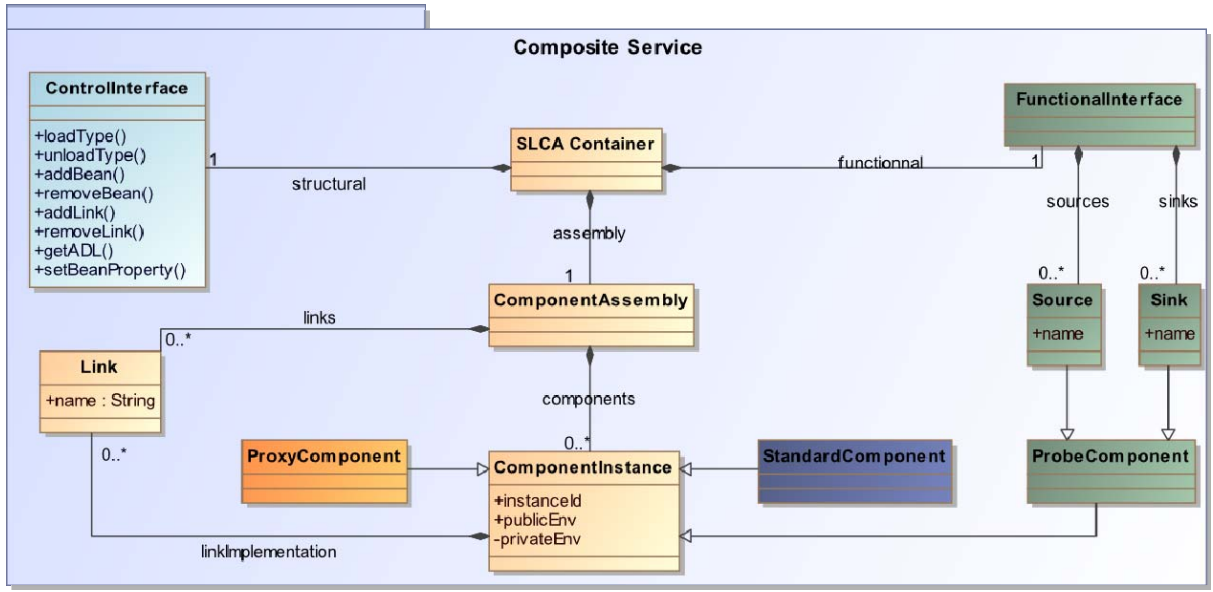

Fig. 4 SLCA Meta-model: interfaces of composite services.

Two types of probe components exist (Fig. 4): sinks, which add a method to the composite service interface, and which, in the internal component assembly, has only an output port. The invocation of the method from the service interface thus emits an event in the component assembly. The second type of probe is the source, which adds an event to the composite service interface, and has only an input port. The invocation of the method from the component interface thus emits a Web service event.

The control interface addresses dynamic modifications of the internal component assembly. It provides methods for adding or removing component instances, types, or bindings, and also to get information about the assembly. Therefore, another client, which can be a composite service using a proxy component for this service interface, can act on the structure of an other composite service. The structural adaptation of composite services and applications is thus possible in the model, by its own entities.

Colors of the UML diagram of Fig. 4 match those of Fig. 2 to make the reading easier. Proxy components allow services of the environment to be used in the composite service, while probe components allow new services to be added to the environment, which can eventually be used by other composite services. The concept of distributed hierarchy is then introduced, through the service layer.

In the first example, we were able to discover and access devices and services of a room in a homogeneous way. The service orchestration with lightweight components then enabled the creation of dynamic applications in the room. This third step, service composition, allows created applications to be reused as a part of new applications. Composite services representing a room export their functionalities and are reused by composite services of a floor, or of a building.

## 4. Experimentation and Validation

### 4.1. Implementation of SLCA Model

The SLCA model has been projected into an implementation called SharpWComp 2.0, which was deposed as copyrighted software in France, used and developed in three programs of the French National Research Agency (ANR). The first explored self-adaptation of software applications to assist people with disabilities. It is creating interaction devices, so they are adapted to profiles of reduced mobility people, and self-adapting to variations of the profile in time. The second project adds contracts inside composite services, like binding the execution time of a service, or catching execution points of an application to add some actions. The aim of the last one is to be able to provide continuous services to a user, with mechanisms of self adaptation of composite services.

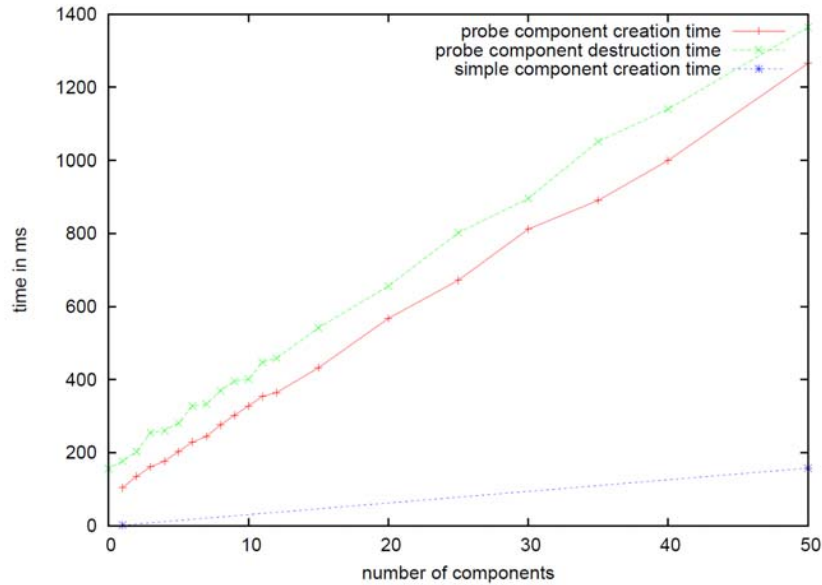

Fig. 5 Component creation and destruction time measures.

**Dynamicity of composition.** The important point about SLCA composite services is that they are fully dynamic. We are able to create composite services, by visual composition of components or textual commands at runtime, which enables a fast application prototyping [28]. But what is valuable now, is that if a service becomes unavailable whilst being in use, we can modify the composite service's internal logic without recompiling the assembly or restarting it.

**Web Services Implementation**. The control interface of the composite service allows us to use several tools to modify dynamically its internal assembly of components, allowing to load, instantiate, destroy components and links. To implement this control interface (and also for the functional interface), we choose the UPnP protocol. Another current good choice could be DPWS as a replacement of the UPnP protocol. These protocols provide a Web Service approach for physical or software objects with the dynamicity, distributivity, autonomy, interoperability preoccupation. Thus, the composite service becomes then completely automatically dynamic and adapting to simple cases environment changes.

### 4.2. Measures and Validation

Service composition in pervasive computing needs to be reactive to take into account changes of the infrastructure quickly and adapt to users' needs. We measured time of creation and destruction of components in a composite service implemented with the previously mentioned SharpWComp 2.0 (Fig. 5).

The creation time of basic components, as well as proxy components, is constant, around 3ms. Therefore, to create n components, $3 \times n$ ms are needed. The removal of such components couldn't have been measured, because we are in a managed memory environment. This is equivalent to dereference the instance of the component, and remove it from the container's list, which was too fast to be measured. Link creation and destruction time are also too simple operations and could not be measured. These measures correspond to the Lightweight Component Architecture (LCA) described in 3.2. For probe components, which in SharpWComp 2.0 rely on Intel's C# UPnP stack, the creation and destruction time are more important. This is due to the fact that when changing the service interface of a composite service, service advertisements are sent to inform that the previous interface is no longer valid, and then they are reissued with the new interface. In UPnP, an advertisement has to be made for each existing service, so if we consider that a probe component creates a service, every new probe will correspond to sending one more message each time. This is why adding the fortieth probe will take nearly one second: announcing thirty-nine service destruction, and announcing forty new services. Of course, this can be optimized. A service can be published only when all its adaptation is complete, reducing component instantiation time.

The time of generation of proxy component is a important factor in our model. We measured it for a standard light device, containing ten methods divided into two services and two evented variables: the average value is 140.6ms. Thus, the time elapsed from the appearance of a service on the infrastructure to the adaptation of a composite service can be calculated. It will be a sum of the proxy component generation time (140.6ms), the component instantiation time (3ms), the adaptation of the composite service time, depending on how many new components are created, especially probe components and their number in the former assembly. Communications on the service layer, which will trigger all these many operations, must be taken into account and can be costly, depending on the average round-trip between hosts. Finally, the whole idea is that the complete adaptation process time has to be low enough to permit application adaptation to occur when needed, without making the application unusable. Consequently, the rate of infrastructure events must not be too close from the adaptation speed of composite service, which would lead them to spend their time adapting and not executing.

## 5. Conclusion

We have demonstrated that SOA (Service oriented Architecture) and its numerous principles are well adapted for pervasive computing. We have detailed and illustrated, in our model called SLCA, various additional principles to meet completely pervasive software constraints like software infrastructure based on services for devices, local orchestrations based on lightweight component architecture and finally encapsulation of such orchestrations into composite services to address distributed composition of services. With regards to performances measures, we can easily distinguish the negligible delays due to lightweight components handling in comparison with the delays due to the network stack (with UPnP in our measures, about 2%). This result reinforce the interest of lightweight composition of services for pervasive computing, where the complete adaptation process time has to be low enough to permit application adaptation to occur when needed, without making the application unusable. Our future works focus on crosscutting modularity, to facilitate incremental evolution of applications, and implementing replicable modification schemes on a large number of services. The Aspect paradigm well known in the object oriented programming field [29], is now widely applied to other architectural paradigms (AO4BPEL for orchestrations [30], FAC for components [31], and so on). Our SLCA model takes the same direction, evolving toward an approach using the concept of Aspect of Assembly (AA). It allows crosscutting evolutions and adaptations of the distributed composition of services for devices.

## Acknowledgments

This work is supported by the French ANR Research Program VERSO in the project *ANR-08-VERS-005* called CONTINUUM.

**IJCSI**

**Jean-Yves Tigli** got his PhD degree in computer science from the University of Nice Sophia Antipolis, in 1996, on software engineering for intelligent robotics systems. He participated in various European projects between 1998 and 2002 (in ESPRIT and MAST European research programs). He's Associate Professor in Computer Science at the Engineering School of Technology of the University of Nice – Sophia Antipolis, France. He's currently managing and leading a project called "Continuum" supported by the French national research agency (ANR) to address the challenge of service continuity in dynamic pervasive environments involving various French universities and international companies.

**Stéphane Lavirotte** got his PhD degree in computer science from the University of Nice – Sophia Antipolis and INRIA, in 2000, on software for document Analysis and Recognition. He participated in various European projects between 1997 and 2004 (in ESPRIT, IST European research programs).
He is Associate Professor in Computer Science at the IUFM of the University of Nice – Sophia Antipolis, France.

**Gaëtan Rey** got his PhD degree in computer science from the University of Joseph Fourrier at Grenoble, in 2005, on context-aware computing. During 2005-2006, he spent one year in the System Research Group of the University College of Dublin,UK. He's Associate Professor in Computer Science in the Institute of Technology of the University of Nice – Sophia Antipolis, France.

**Vincent Hourdin** is preparing his PhD thesis on context-based security in SOA for pervasive computing at the University of Nice – Sophia Antipolis, supervised by J.-Y. Tigli and Michel Riveill. He's also software research engineer for MobileGov, an IT security software editor in Sophia Antipolis, France.

**Michel Riveill** got his PhD degree in computer science from the National Polytechnic Institute of Grenoble, in 1987, on distributed software. He obtained "Habilitation à Diriger les Recherches" in 1993, from the same institute. He's full Professor in Computer Science at the Engineering School of Technology of the University of Nice – Sophia Antipolis, France. He's leading the computer science department of the engineering school and the software engineering department of the computer science laboratory of the University of Nice - Sophia Antipolis and CNRS.